\documentstyle[12pt,epsf]{article}
\newcommand{\op}[1]{\mathop{\rm #1}\nolimits}

\newcommand{\scrarr}[2]{{\ba{l}\scs{#1}\\[-10pt]\scs{#2}\ea}}
\newcommand{\be}{\begin{equation}}
\newcommand{\ee}{\end{equation}}
\newcommand{\ba}{\begin{array}}
\newcommand{\ea}{\end{array}}
\newcommand{\bea}{\begin{eqnarray}}
\newcommand{\eea}{\end{eqnarray}}
\newcommand{\by}{\begin{eqnarray*}}
\newcommand{\ey}{\end{eqnarray*}}

\newcommand{\ve}{\varepsilon}
\newcommand{\p}{\partial}
\newcommand{\ra}{\rightarrow}
\newcommand{\La}{\Lambda}

\newcommand{\al}{\alpha}

\newcommand{\si}{\sigma}

\newcommand{\f}{\frac}
\newcommand{\ti}{\tilde}
\newcommand{\pr}{\prime}
\newcommand{\ct}{\cite}

\newcommand{\Ga}{\Gamma}
\newcommand{\scs}{\scriptstyle}
\newcommand{\ds}{\displaystyle}

\newcommand{\ts}{\times}
\newcommand{\ul}{\underline}
\newcommand{\nn}{\nonumber}

\begin{document}
\begin{center}
{\LARGE Magnetic catalysis in a $P$-even, \\
 chiral-invariant 3-dimensional model \\
 with four-fermion interaction\\[5mm]

}{\em
 V.Ch. Zhukovsky$^\dagger$, K.G. Klimenko$^{\ast\ddagger}$ and V.V. Khudyakov\\[4mm]

}{\footnotesize
Moscow State University , 117234, Moscow, Russia\\
 $^\ast\;$Institute of High Energy Physics, 142284, Protvino, Russia\\
 E-mail:~~$^\dagger\;$th180@phys.msu.su;~~$^\ddagger\;$kklim@mx.ihep.
 su\\
}

\vspace*{1cm}
\parbox{150mm}{\small
The influence of an external constant and homogeneous magnetic field
$H$ on the phase structure of the
$P$-symmetric, chiral invariant 3-dimensional field theory model
with two four-fermion interaction structures is considered.
An arbitrary small (nonzero) magnetic field is shown to induce
spontaneous violation of the initial symmetry (magnetic catalysis).
Moreover, vacuum of the model at $H\neq 0$ can be either
$P$-symmetric
or chiral invariant, depending on the values of the coupling
constants.
} \end{center} \vspace*{4mm}

\section{Introduction}

Magnetic catalysis is the phenomenon of dynamical symmetry breaking
induced by external magnetic or chromomagnetic fields.
This kind of influence of an external magnetic field
$H$ was for the first time observed in the study of the
(2+1)-dimensional chiral invariant Gross--Neveu model~\cite{gn}. In
that case, an arbitrary weak external magnetic field lead to
dynamical chiral symmetry breaking (D$\chi$SB) even for an
arbitrary small coupling constant~\cite{1,krive}.
This phenomenon is explained by the effective reduction
of space-time dimensions in the external magnetic field and,
correspondingly, by strengthening the role of the infrared divergences
in the vacuum reorganization~\cite{3}. Later, it has been shown that
spontaneous chiral symmetry breaking can be induced by an external
chromomagnetic field as well~\cite{2,obzor,ebert}. Moreover, basing
upon the study of a number of field theories, it has been argued that
magnetic catalysis of D$\chi$SB may have a universal, i.e. model
independent, character~\cite{w}. In a recent paper~\cite{liu}, a $P$ symmetric
3-dimensional Gross--Neveu model has been studied, and the external magnetic
field proved to be a catalyst for spontaneous symmetry breaking again, this
time for $P$ parity breaking. Magnetic catalysis has already found its
applications in cosmology and astrophysics~\cite{10}, and in constructing
the theory of high-temperature superconductivity~\cite{liu,11}. It
is evident that this phenomenon may be of crucial importance for
various processes in the presence of an external magnetic field, where the
dynamical symmetry breaking takes place, i.e. in the
elementary-particle physics, condensed-matter physics, physics of
neutron stars etc.\footnote{Possible applications of the
magnetic catalysis were also discussed in recent
publications~\cite{g}.}

As it is well known~\cite{d}, the high temperature superconductivity
is observed in antiferromagnetic materials like La$_2$CuO$_4$,
where conduction electrons are concentrated in planes formed by
atoms of Cu and O. In recent experimental studies of this
phenomenon~\cite{kr}, it was found out that at temperatures much
lower than the temperature of transition to the superconducting state,
the external magnetic field induces a parity-breaking phase transition.
Thus, the problem of studying magnetic catalysis in
high-temperature superconductivity becomes quite urgent. The above
arguments are first of all applicable to (2+1)-dimensional
field-theoretical models of high-temperature
superconductivity.

In the present paper, we consider the influence of an external magnetic
field on one of these 3d models with a four-fermion interaction
Lagrangian of the form
\be
\label{eq.1}
L=\sum_{a=1}^N\bar\psi_a i\hat\p\psi_a+\f{G_1}{2N}
(\sum_{a=1}^N\bar\psi_a\psi_a)^2
+\f{G_2}{2N}(\sum_{a=1}^N\bar\psi_a\tau\psi_a)^2,
\ee
where $\hat\p\equiv\Ga^\mu\p_\mu\,$, fields $\psi_a$ are transformed
according to the fundamental representation of the $U(N)$ group
introduced in order to employ the nonperturbative $1/N$-expansion
method. Moreover, $\psi_a$ is a four-com\-po\-nent Dirac spinor for
every value of $a=1,2,...,N$ (corresponding indices are omitted). The
action with the Lagrangian (\ref{eq.1}) is invariant under two discrete
transformations, one of them being parity ($P$), and the other is
chiral symmetry $\Gamma_5$ (these operations, as well as the matrix
$\tau$ are defined in the Appendix, where the algebra of $\Gamma^\mu$
matrices for the spinor reducible 4-di\-men\-sio\-nal representation
of the Lorentz group is also presented).

The model (\ref{eq.1}) at $H=0$ has already been studied in
papers~\cite{sem}, and a corresponding gauge version of the model was
considered in ~\cite{carena}. This interest is mainly justified by the
fact, that in a number of lattice models, describing dynamic
effects in quantum antiferromagnetics in two spatial dimensions (e.g.,
high temperature superconductivity), going to continuum limit renders
theories of the type (\ref{eq.1}).

In contrast to simplest Gross-Neveu models, there are two different
structures with four-fermion interaction in the Lagrangian (\ref{eq.1}).
Hence, two different fermion condensates can exist in the theory: one
of them, $\langle\bar\psi\psi\rangle$, is the order parameter for
chiral symmetry, and the other one, $\langle\bar\psi\tau\psi\rangle$,
characterizes spontaneous breaking of parity. The condensate that
provides for the lowest vacuum energy corresponds to the symmetry that
is broken. The above arguments are fully applicable to the case of
nonzero field $H\neq 0$. Therefore, the magnetic catalysis
effects in the model (\ref{eq.1}) and in simplest Gross-Neveu models
should be qualitatively different.

Our main goal is to
study characteristics of the magnetic catalysis in the
$P\times\Gamma_5$-invariant model (\ref{eq.1}). At first, we
are going to find out whether spontaneous symmetry breaking of the initial
symmetry is induced by an external magnetic field. If this is the
case, the question is whether both $P$ and $\Gamma_5$ symmetries are
broken simultaneously, or the vacuum of the model retains symmetry with
respect to only one of the two discrete transformations. If the latter
possibility is realized, then a question arises how the residual
symmetry of the model depends on the values of coupling constants
$G_{1,2}$, etc. We demonstrate, that under certain conditions in the
framework of the field theory (\ref{eq.1}), a parity breaking
phase transition induced by an external magnetic field is possible.

\section{Phase structure of the model at $H=0$}

Before considering the influence of the nonzero external magnetic
field $H$ on the vacuum of the model (\ref{eq.1}), its phase structure
will be studied at $H=0$. To this end, the auxiliary Lagrangian \be
\ti L=\bar\psi
i\hat\p\psi+\si_1(\bar\psi\psi)+\si_2(\bar\psi\tau\psi)
+\f{N\si_1^2}{2G_1}+\f{N\si_2^2}{2G_2}
\label{eq.2}
\ee
is introduced, where $\si_{1,2}$ are auxiliary boson fields, and
summation over indices $a$ of the auxiliary group $U(N)$ is implicitly
performed here and in what follows. The field theories (\ref{eq.1}) and
(\ref{eq.2}) are equivalent,
since by means of the equations of motion the fields $\si_{1,2}$ can be
excluded from (\ref{eq.2}) and the Lagrangian (\ref{eq.1}) obtained. It is
easily shown that the auxiliary fields are transformed with respect to
discrete symmetries $P$ and $\Ga_5$ in the following way:
\be\ba{l}
P:~\si_1(t,x,y)\ra\si_1(t,-x,y);~~\si_2(t,x,y)\ra-\si_2(t,-x,y); \\
\Gamma_5:~\si_1(t,x,y)\ra-\si_1(t,x,y);
~~\si_2(t,x,y)\ra\si_2(t,x,y),
\label{eq.3}
\ea\ee
i.e. $\si_{1}$ is a scalar field, and $\si_{2}$ is pseudoscalar one.
Starting from the Lagrangian (\ref{eq.2}), one can find the effective
action of the theory, which has the following form in the one-loop
approximation (i.e. in the leading order of the $1/N$ expansion):
\be
S_{\rm eff}(\si_1,\si_2)=-N\int d^3x\left
(\f{\si_1^2}{2G_1}+\f{\si_2^2}{2G_2}\right )
-i\mbox{Sp}\ln\Delta,
\label{eq.4}
\ee
where $\Delta = i\hat\p+\si_1+\si_2\tau$. Here, the fields
$\si_{1,2}$ are functions of space-time points. To obtain the
effective potential of the model, the definition has to be used
\be \left.
V_0(\si_1,\si_2)\int d^3x=-S_{\rm eff}(\si_1,\si_2)
\right|_{\si_1,\si_2 =\op{const}},
\label{eq.5}
\ee
where boson fields are assumed to be independent of coordinates. With
the help of (\ref{eq.4}) and (\ref{eq.5}), the following expression for the
effective potential of the model (\ref{eq.1}) can be found in the leading
order of the $1/N$ expansion:
\be
V_0(\si_1,\si_2)=\f{N\si_1^2}{2G_1}+\f{N\si_2^2}{2G_2}
-N\sum_{k=1}^2\int\f{d^3p}{(2\pi)^3}\ln(p^2+M^2_k),
\label{eq.6}
\ee
where $M_{1,2}=|\si_2\pm\si_1|$, and integration is performed over Euclidean
momentum. Integration in (\ref{eq.6}) over the domain
$0\leq p^2\leq \La^2$ yields:
\be
V_0(\si_1,\si_2)=N\sum_{k=1}^2\left[\f{\si_k^2}{2}\left
(\f{1}{G_k}-\f{2\La}{\pi^2}\right) +\f{M_k^3}{6\pi}\right].
\label{eq.7}
\ee
Now, in order to exclude the cutoff parameter $\La$ from (\ref{eq.7}), we
introduce the renormalized coupling constants with the help of the
following normalization conditions:
\bea \left.
\f1N\f{\p^2V_0}{(\p\si_1)^2}\right|_\scrarr{\si_1=\nu_1}{\si_2=0}
 = \f{1}{G_1}-\f{2\La}{\pi^2}+\f{2\nu_1}{\pi}
\equiv\f{1}{g_1(\nu_1)},\nn \\
\left.\f1N\f{\p^2V_0}{(\p\si_2)^2}\right|_\scrarr{\si_1=0}{\si_2= \nu_2}
 = \f{1}{G_2}-\f{2\La}{\pi^2}+\f{2\nu_2}{\pi}
\equiv\f{1}{g_2(\nu_2)}.
\label{eq.8}
\eea
Relations (\ref{eq.8}) enable us to write the effective potential in terms
of the ultraviolet finite quantities
\be
V_0(\si_1,\si_2)=N\sum_{k=1}^2\left[\f{\si_k^2}{2g_k}
+\f{M_k^3}{6\pi}\right],
\label{eq.9}
\ee
where ($i=1,2$):
\be
\f1{g_i}=\f{1}{g_i(\nu_i)}-\f{2\nu_i}{\pi}=\f{1}{G_i}-\f{1}{G_c}
\label{eq.10}
\ee
and $G_c=\pi^2/(2\La)$. We remind that the coupling constants
$G_i$ ($g_i(\nu_i)$)
depend (do not depend) on the cutoff parameter $\La$ and do not depend
(depend) on the normalization masses $\nu_i$. By virtue of this,
(\ref{eq.10}) leads us to the conclusion that the constants $g_i$
depend neither on $\nu_i$, nor on $\La$, i.e. the effective potential
$V_0$ is a renormalization invariant finite quantity (i.e., without
ultraviolet divergences).

Thus, we have indeed demonstrated that the model (\ref{eq.1}) is
renormalizable in the leading order of the $1/N$ expansion. The complete
proof of its renormalizability is not intended in the
present paper. However, we believe this to be true,
relying on the results of the review paper~\ct{ros}, where the simplest
3-dimensional models with four-fermion interaction were proved to be
renormalizable in the framework of the nonperturbative $1/N$ expansion
method.

It is well known that the phase structure of any theory is to a great extent
determined by the symmetry of the global minimum of its effective
potential. In order to avoid cumbersome formulas, we omit intermediate
calculations, related to the study of the absolute minimum of
the potential $V_0(\si_1,\si_2)$ (\ref{eq.9}) and present here the final
result -- the phase
portrait of the model (\ref{eq.1}), depicted in Fig.~1 (similar analysis of
the effective potentials for a whole series of the theories with the
four-fermion interaction is carried out in~\ct{obzor,klim}, and, if
required, it can be
easily performed here). In this figure, the plane $(g_1,g_2)$ of
coupling constants (\ref{eq.10}) is divided into five regions, in each of
which one of the possible phases of the model is realized: A, B, C. In the
phase A, where $g_{1,2}>0$, corresponding to small values of the
coupling constants $G_{1,2}<G_c$, the global minimum of the function
$V_0(\si_1,\si_2)$ is at the origin $(\si_1=\si_2=0)$, and hence, the ground
state of the system in this phase is $P$ and $\Ga_5$ symmetric. In phases B
and C, where at least one of the coupling constants $G_i$ is greater than
$G_c$, the
point of the potential global minimum is correspondingly at
$\si_1=0,\;\si_2=-\pi/g_2$ and $\si_1=-\pi/g_1,\;\si_2=0$. Hence, with
the aid of (\ref{eq.3}), one can see that the vacuum is $\Ga_5$ invariant in
the phase B, and the parity is spontaneously broken. The ground state of
the phase C of the model is $P$ invariant, however the chiral symmetry is
spontaneously broken in this case.

\section{Effective potential at $H\ne0$}

Before starting to investigate the vacuum of the 3-dimensional model (\ref{eq.1})
in the external constant magnetic field, let us work out an expression for
the effective potential at $H\neq 0.$ The Lagrangian of the model in
terms
of auxiliary scalar fields $\si_{1,2}(x)$ takes the form
\be
\ti L_H=\bar\psi(i \hat \partial+e\hat A)\psi
+\si_1(\bar\psi\psi)+\si_2(\bar\psi\tau\psi)
+\f{N\si_1^2}{2G_1}+\f{N\si_2^2}{2G_2},
\label{eq.11}
\ee
where $\hat A\equiv A_{\mu}\Gamma^{\mu},\;e$ is the (positive)
charge of fermions, and the vector-potential of the constant external
magnetic field $H$ has the form $A_{0,1}=0,\;A_2=x_1H.$ Using the
standard
technique of the $1/N$ expansion~\cite{gn}, one obtains, in the
leading
$1/N$ order, the
following expression for the effective potential:
\be
V_{\rm eff}(\si_1,\si_2)=\f{N\si_1^2}{2G_1}+\f{N\si_2^2}{2G_2}+
\f{iN}{v}\mbox{Sp}\ln\Delta_H.
\label{eq.12}
\ee
Let us note, that here $\Delta_H=i \hat \partial+e\hat
A+\si_1+\tau\si_2$
is an operator that acts only in the spinor and coordinate spaces,
$v=\int
d^3x$,
and fields $\si_i$ do not depend on space-time points.

First, let us turn to the causal Green's function of the operator
$\Delta_H$. It can be presented in the following form:
\bea
(\Delta_H^{-1})_{\al\beta}(\vec x,t;\vec x^{\prime},t^{\pr})&=&
-i\theta(t-t^{\prime})
\sum_{\{n\}}~\psi^{(+)}_{\{n\}\al}(\vec x,t)
\bar\psi^{(+)}_{\{n\}\beta}(\vec x^{\prime},t^{\prime})+\nn \\
&&+i\theta(t^{\prime}-t)
\sum_{\{n\}}~\psi^{(-)}_{\{n\}\al}(\vec x,t)
\bar\psi^{(-)}_{\{n\}\beta}(\vec x^{\prime},t^{\prime}).
\label{eq.13}
\eea
Here $\{n\}=(i,n,k),$ where $i=1,2;\;n=0,1,2,...;\;k$ is a real
number
$-\infty<k<\infty.$ Moreover, $\psi^{(\pm)}_{\{n\}}$ are
positive-~and
negative-frequency orthogonal normalized solutions of the Dirac
equation
$\Delta_H\psi=0$, which in the case $\si_1+\si_2>0,\;\si_2<\si_1$
have the form (T is a symbol of transposition):
\be\ba{l}
\psi^{(\pm)T}_{1,n,k}(\vec x,t)= \exp(\mp i\ve_nt+ikx_2)
\times\\[1mm]
\qquad\qquad\ds
 \times\left(\mp\sqrt{\f{\ve_n\mp(\si_1+\si_2)}{4\pi\ve_n}}
h_{n,k}(x_1),\sqrt{\f{\ve_n\pm(\si_1+\si_2)}{4\pi\ve_n}}
h_{n-1,k}(x_1),0,0\right), \\[4mm]
\psi^{(\pm)T}_{2,n,k}(\vec x,t)= \exp(\mp i\ti\ve_nt+ikx_2) \times
\\[1mm]
\qquad\qquad\ds
 \times\left(0,0,\sqrt{\f{\ti\ve_n\pm(\si_1-\si_2)}{4\pi\ti\ve_n}}
h_{n,k}(x_1),\mp\sqrt{\f{\ti\ve_n\pm(\si_2-\si_1)}{4\pi\ti\ve_n}}
h_{n-1,k}(x_1)\right),
\label{eq.14}
\ea\ee
where
\be
h_{n,k}(x_1)=\f{(eH)^{1/4}}{(2^nn!\sqrt{\pi})^{1/2}}\exp(-\xi^2/2)H_n
(\xi),
\label{eq.15}
\ee
 $\ve_n=\sqrt{(\si_1+\si_2)^2+2eHn}$,
$\ti\ve_n=\sqrt{(\si_1-\si_2)^2+2eHn},\;$ $H_n(\xi)$ are the Hermite
polynomials,
$\xi=\sqrt{eH}\left[x_1-k/(eH)\right].$
Functions (\ref{eq.15}) satisfy conditions
\be
\int dx_1~h^2_{n,k}(x_1)=\f{1}{eH}\int dk~h^2_{n,k}(x_1)=1.
\label{eq.16}
\ee
Moreover, in formulas (\ref{eq.14}), it is assumed that
$h_{-1,k}(x_1)\equiv 0.$

Now we have obtained everything necessary for calculating the value
$A(\si_1,\si_2)\equiv iv^{-1}\mbox{Sp}\ln\Delta_H$. It is evident,
that
\be
\f{\p A(\si_1,\si_2)}{\p\si_1} =iv^{-1}\mbox{Sp}\Delta_H^{-1}
\equiv iv^{-1}\int d^3x\sum_{\al=1}^4
(\Delta_H^{-1})_{\al\al}(\vec x,t;\vec x,t).
\label{eq.17}
\ee
Taking formulas (\ref{eq.13})--(\ref{eq.16}) into account, after
rather tedious calculations, one obtains from this expression
(similar
calculations were performed in the case of the simplest Gross--Neveu
model in ~\cite{obzor}): \bea \f{\p A(\si_1,\si_2)}{\p\si_1}&=&
-\f{(\si_1+\si_2)eH}{4\pi^{3/2}}
\int\limits_{0}^{\infty} \f{ds}{\sqrt{s}}
 \exp (-s(\si_1+\si_2)^2)\op{cth}(eHs)\nn\\
&&-\f{(\si_1-\si_2)eH}{4\pi^{3/2}}
\int\limits_{0}^{\infty} \f{ds}{\sqrt{s}} \exp
(-s(\si_1-\si_2)^2)\op{cth}(eHs).
\label{eq.18}
\eea
Integrating both sides of this equality over $\si_1$ within the
limits $\si_1$ and $\infty$, one obtains
\be
A(\si_1,\si_2)- A(\infty,\si_2)=
\f{eH}{8\pi^{3/2}}\sum_{i=1}^2
\int\limits_{0}^{\infty} \f{ds}{s^{3/2}} \exp(-sM_i^2)\op{cth}(eHs),
\label{eq.19}
\ee
where expressions for $M_i$ are presented in (\ref{eq.6}). Note,
that formula (\ref{eq.19}) was obtained with the assumption that
$\si_2>-\si_1$, $\si_2<\si_1$. However, one can directly prove that
(\ref{eq.19}) is also valid for arbitrary relation between
$\si_1$ and $\si_2$.

The value of $\p A(\si_1,\si_2)/\p\si_2$ can be found in a
similar way. One can integrate it over $\si_2$ within the
limits $\si_2$ and $\infty$, and then the expression obtained can be
compared with (\ref{eq.19}). As a result, we obtain the equality
 $A(\infty,\si_2)=A(\si_1,\infty)$, which leads to the conclusion
 that,
due to mutual independence of variables $\si_1$ and $\si_2$, the
value
of $A(\infty,\si_2)$ does not depend on $\si_2$, i.e. it is constant.

Taking this into account, due to equation (\ref{eq.19}),
one obtains from (\ref{eq.12}) the following expression for the
effective
potential (up to an inessential constant):
\be
 V_{\rm eff}(\si_1,\si_2)=\f{N\si_1^2}{2G_1}+\f{N\si_2^2}{2G_2}+
 N\sum_{k=1}^2 \f{eH}{8\pi^{3/2}}
 \int\limits_{0}^{\infty} \f{ds}{s^{3/2}} \exp(-sM_k^2)\op{cth}(eHs).
\label{eq.20}
\ee
This expression evidently contains UV divergent integrals.
Upon some transformations in (\ref{eq.20}),
this divergence can be localized in the part of the effective
potential
that corresponds to $H=0$ contribution $V_0(\si_1,\si_2)$:
\be
 V_{\rm eff}(\si_1,\si_2)=V_0(\si_1,\si_2)+N\sum_{k=1}^2
 \f{eH}{8\pi^{3/2}} \int\limits_{0}^{\infty} \f{ds}{s^{3/2}}
 \exp(-sM_k^2)
 \left[\op{cth}(eHs)-\f{1}{eHs}\right].
\label{eq.po}
\ee
Here
\be
 V_0(\si_1,\si_2)=\f{N\si_1^2}{2G_1}+\f{N\si_2^2}{2G_2}+
 N\sum_{k=1}^2\f1{8\pi^{3/2}}
 \int\limits_{0}^{\infty} \f{ds}{s^{5/2}} \exp (-sM_k^2).
\label{eq.22}
\ee
It is easily seen that only the first term in (\ref{eq.po}), defined
in
(\ref{eq.22}), is singular. Up to a certain additive constant, it is
equal
to potential (\ref{eq.6}).
Upon renormalization, it evidently takes the form of (\ref{eq.9}).
 The following expression for $V_{\rm eff}$, resulting from
 (\ref{eq.po})
after integration over $s$ (see~\cite{prud}), will also be used:
\bea
 V_{\rm eff}(\si_1,\si_2)=
 N\sum_{k=1}^2\left[ \f{\si_k^2}{2g_k} +
 \f{eHM_k}{4\pi}-\f{(2eH)^{3/2}}{4\pi}\zeta\left(-\f{1}{2},
 \f{M_k^2}{2eH}\right) \right],
\label{eq.pot}
\eea
where $\zeta (s,x)$ is the generalized Riemann
Zeta-function~\cite{yit}.

\section{The phenomenon of magnetic catalysis}

Now, let us consider spontaneous symmetry breaking in
the initial model (\ref{eq.1}).
To this end, properties of the
global minimum of the potential (\ref{eq.pot}) under
symmetry transformation (\ref{eq.3}) have to be investigated.
 In this case, positive values of parameters $g_{1,2}\,$, i.e.,
correspondingly,
the region of small values of the bare coupling constants
$G_{1,2}\,$,
 where the
$P\times\Ga_5$-symmetric phase of the model exists at $H=0,$ are of
 special interest. The question is what
 kind of symmetry the vacuum possesses at $H\neq 0$ in the case under
consideration, i.e. at $g_1>0$, $g_2>0$. To find an answer to this
question,
it is useful to consider a simplified situation.

\ul{\bf Special case of model (\ref{eq.1}).}
First, we illustrate the magnetic catalysis phenomenon by
considering the example of a simplest model that follows from
(\ref{eq.1})
when $G_1\equiv G$, $G_2=0$.
The corresponding effective potential $V_{Hg}(\si)$ can be obtained
from (\ref{eq.pot}), assuming $\si_2 \equiv 0$, and introducing
the notations
$\si_1\equiv\si,~g_1\equiv g$:
\be
\label{eq.k1}
V_{Hg}(\si)=\f{N\si^2}{2g}+\f{NeH|\si|}{2\pi}
-\f{N(2eH)^{3/2}}{2\pi}\zeta\left(-\f{1}{2},\f{\si^2}{2eH}\right).
\ee
It is evident that function $V_{Hg}(\si)$ is symmetric under
the transformation $\si\to -\si$. Hence, in order to find its global
minimum, it is sufficient to study
$V_{Hg}(\si)$ only in the domain
$\si\in [0,\infty)$. Here, the stationary equation has the form
\be
\label{eq.k2}
\f{\p V_{Hg}(\si)}{\p\si}=\f{N\si}{g}+\f{NeH}{2\pi}-
\f{N\si\sqrt{2eH}}{2\pi}\zeta\left(\f{1}{2},\f{\si^2}{2eH}\right)=0.
\ee
This can be obtained with account for the fact that
$$d\zeta (s,x)/dx=-s\zeta (s+1,x).$$
When $\si\ra 0$, the following expansion for the $\zeta$-function is
known \cite{yit}: \be \label{eq.k3}
\zeta\left(\f{1}{2},\f{\si^2}{2eH}\right)=\f{\sqrt{2eH}}{|\si|}
+\mbox{const}+o(|\si|/\sqrt{2eH}).
\ee
Substitution of (\ref{eq.k3}) into a LHS of (\ref{eq.k2}) gives
\be
\left.\f{\p V_{Hg}(\si)}{\p\si}\right|_{\,\si\to 0_+}=-\f{NeH}{2\pi},
\label{eq.k100}
\ee
i.e. the point $\si=0$ is not a solution of the stationary
equation
(\ref{eq.k2}). Moreover, since $V_{Hg}(\si)=V_{Hg}(-\si)$, two more
consequences follow from (\ref{eq.k100}): 1. The point $\si=0$ is a
local maximum of
potential
(\ref{eq.k1}). 2. The first derivative of function $V_{Hg}(\si)$
does not exist at the point $\si=0$. (At the same time, the effective
potential for $H=0$ is a
differentiable function in the whole axis $\si=0$.) Consequence n.1
and the
fact that $\lim_{|\si|\to\infty}V_{Hg}(\si)=+\infty$, imply that
there
exists
a point $\si_0(H,g),$ where the potential $V_{Hg}(\si)$ has a
nontrivial,
i.e. nonvanishing global minimum. This means that chiral symmetry of
the
simplest model in question is necessarily spontaneously broken in the
presence of an arbitrary weak external magnetic field, when the
coupling
constant $G$ is arbitrarily small (constant $g$ is positive). Thus,
dynamical breaking of chiral invariance, induced by an external
magnetic
field, takes place (magnetic catalysis). More detailed analysis of
the
phase
 structure of the simplest Gross-Neveu model in an external magnetic
 field
can be found in \ct{1,3,obzor,eliz}, where the influence of
temperature,
chemical potential and nonzero space-time curvature on the magnetic
catalysis
is also studied.\footnote{Influence of such factors as temperature,
chemical
potential, nontrivial metrics and topology of the space (disregarding
magnetic field) on the ground state structure in the simplest
(2+1)-dimensional Gross--Neveu model is considered in~\cite{vitale}.}

In what follows, more detailed information on the properties of the
effective potential (\ref{eq.k1}) will be needed.
Assume that $g>0$
and $eHg^2\ll 1$, i.e. the magnetic field is weak. In this case, with
the
use of expansion (\ref{eq.k3}) in equation (\ref{eq.k2}), one can
easily
obtain the following asymptotic behavior of the global minimum point
$\si_0(H,g)$ of potential $V_{Hg}(\si)$ at $eH\to 0$, being a
solution of
the stationary equation (\ref{eq.k2}):
\be
\label{eq.k4}
\si_0(H,g)=egH/(2\pi)+\ldots\, .
\ee

We now consider the quantity $V_{Hg}(\si_0(H,g))$.
It is evident that for arbitrary values of $H$ and $g$ the following
relation holds:
\be
\label{eq.k5}
\f{dV_{Hg}(\si_0(H,g))}{dg}=\left.\f{\p\si_0(H,g)}{\p g}
 \f{\p V_{Hg}(\si)}{\p\si}\right|_{\,\si=\si_0(H,g)}
 +\left.\f{\p V_{Hg}(\si)}{\p g}\right|_{\,\si=\si_0(H,g)}.
\ee
The first term in the right-hand side of the equation vanishes in
virtue of the fact that $\si_0(H,g)$ satisfies the stationary
equation (\ref{eq.k2}). Thus, we have
\be
\label{eq.k6}
\left.\f{dV_{Hg}(\si_0(H,g))}{dg}=\f{\p V_{Hg}(\si)}{\p g}
 \right|_{\,\si=\si_0(H,g)}= -\f{\si_0(H,g)^2}{2g^2}<0.
\ee
It follows from (\ref{eq.k6}) that $V_{Hg}(\si_0(H,g))$ is a
monotonically decreasing function of parameter $g$.

\ul{\bf Magnetic catalysis in the general case.} We now consider
spon\-ta\-ne\-ous sym\-met\-ry brea\-king in model (\ref{eq.1}) under
the
influence of an external magnetic field in the general case, when
two coupling constants $G_{1,2}$ are arbitrary. First of all, we
will demonstrate that the vacuum of the model at $H\neq 0$ is not
$P\times \Ga_5$ symmetric any more, i.e. the point
$\si_{1,2}=0$ does not correspond to the global minimum of the
function
$V_{\rm eff}(\si_1,\si_2)$. To this end, we study first derivatives
of
this function. They have the following form in the domain
$(\si_1+\si_2)>0,$ $\si_k>\si_l$ (here $k\neq l$):
\bea
\f 1N\f {\p V_{\rm
eff}(\si_1,\si_2)}{\p\si_k}&=&\f{\si_k}{g_k}+\f{eH}{2\pi}-
\f{(\si_1+\si_2)\sqrt{2eH}}{4\pi}\zeta\left(\f{1}{2},\f{(\si_1+\si_2)
^2}{2eH
}\right)- \nn \\
&&
-\f{(\si_k-\si_l)\sqrt{2eH}}{4\pi}\zeta\left(\f{1}{2},\f{(\si_1-\si_2
)^2}{2e
H}\right),\nn \\
\f 1N\f {\p V_{\rm eff}(\si_1,\si_2)}{\p\si_l}&=&\f{\si_l}{g_l}-
\f{(\si_1+\si_2)\sqrt{2eH}}{4\pi}\zeta\left(\f{1}{2},\f{(\si_1+\si_2)
^2}{2eH
}\right)+\nn \\
&&
+\f{(\si_k-\si_l)\sqrt{2eH}}{4\pi}\zeta\left(\f{1}{2},\f{(\si_1-\si_2
)^2}{2e
H}\right).
\label{eq.k7}
\eea
If we put $k=2,$ $l=1,$ $\si_1=0$ first, and then $k=1,$ $l=2,$
$\si_2=0$
in
these formulas, one can easily see, with the help of expansion
(\ref{eq.k3}), that
\be
\left.\f{\p V_{\rm eff}}{\p\si_2}\right|_\scrarr{\si_2\to
0_+}{\si_1=0}
 =-\f{NeH}{2\pi}<0;~~~
\left.\f{\p V_{\rm eff}}{\p\si_1}\right|_\scrarr{\si_1\to
0_+}{\si_2=0}
 =-\f{NeH}{2\pi}<0.
\label{eq.k8}
\ee
Relations (\ref{eq.k8}) imply that, when $H\neq 0$, effective
potential
(\ref{eq.pot})
 decreases at the point $\si_{1,2}=0$ in the positive directions of
 axes
$\si_1$ and $\si_2$. Hence, this point can not correspond
to a global minimum of the potential $V_{\rm eff}(\si_1,\si_2)$, and
an
arbitrary weak external magnetic field induces spontaneous breaking
of the
initial $P\times \Ga_5$ symmetry for arbitrary coupling constants
$G_{1,2}$.

Nondifferentiability of functions (\ref{eq.pot}) on the line
$\si_1=\si_2$
is another important consequence of formulas (\ref{eq.k7}). This
results
from
the fact that each of the partial derivatives (\ref{eq.k7}) takes
different values, when approaching this line from above and from
below. Moreover, since the function (\ref{eq.pot}) is symmetric
against
each of the transformations $\si_1\ra -\si_1$, $\si_2\ra -\si_2$, it
is evidently nondifferentiable in the plane of variables
$(\si_1,\si_2)$ on the lines $\si_1=\si_2$ and $\si_1=-\si_2$. We
note
that at $H=0$ the effective potential is differentiable in the whole
plane $(\si_1,\si_2)$. The function (\ref{eq.pot}) is even in each
of its
variables and hence, we will further study its global minimum only
in
the domain $\si_1\geq 0,~\si_2\geq 0$.

Consider the influence of an external magnetic field on the symmetric
phase
of
the model in more detail, i.e. we assume that $g_1$ and $g_2$ are
positive.
It follows from (\ref{eq.10}) that bare coupling constants are
sufficiently
small in this case, i.e. $G_i<G_c= \pi^2/(2\La)$.
As it was shown above, the magnetic field destroys the $P\times
\Ga_5$
symmetry of the ground state of the model, i.e. magnetic field is a
catalyst of the spontaneous breaking of this symmetry.
In order to study magnetic catalysis in more detail, we first
assume that the magnetic field is so small that $eHg^2_1\ll 1$,
$eHg^2_2\ll 1$. Then only two points satisfy the system of
stationary equation
$\p V_{\rm eff}/\p\si_k=0$ (where $k=1,2$):
 i. $(\si_{10},0)$ and ii. $(0,\si_{20})$, where
$\si_{i0}=\si_0(H,g_i)$, and $\si_0(H,g_i)$ are the solutions of
equation
(\ref{eq.k2}), where $g=g_i$. It is evident
that the global minimum of potential $V_{\rm eff}(\si_1,\si_2)$ can
be
situated only in one of these two points on the
$(\si_1,\si_2)$-plane,
where the following relations hold
\be
V_{\rm eff}(\si_{10},0)=V_{Hg_1}(\si_0(H,g_1));~~
V_{\rm eff}(0,\si_{20})=V_{Hg_2}(\si_0(H,g_2)).
\label{eq.k9}
\ee
In the right hand sides of these equations, the minimum value of
function
(\ref{eq.k1}) at $g=g_1$ and $g=g_2$ stands. According to
(\ref{eq.k5}) and
(\ref{eq.k6}), the values of the function
in the right hand sides of these equations decrease monotonically
with
parameter $g$. Therefore, at $g_1<g_2$ ($g_2<g_1$) the effective
potential (\ref{eq.po}) is smaller at the point ii (i) than at the
point i (ii).

The final conclusion about the position of the global minimum of
potential (\ref{eq.pot}) can be drawn only after studying this
function
on the line $L=\{(\si_1,\si_2):$ $~\si_2=\si_1\}$. On this line, as
it
was shown above, the potential is not differentiable, and hence,
there
exists a probability that the global minimum of the potential
$V_{\rm eff}(\si_1,\si_2)$ is situated just here. Consider in more
detail
such a possibility, and to this end, we study the function
(\ref{eq.pot}) on the straight line $L$ where it has the form
\be
V_{\rm eff}(\si,\si)\equiv \ti V(\si)=\f {V_{H\ti g}(2\si)}2-
\f {N\si^2}{2\ti g}+\f {V_{H\ti g}(0)}2.
\label{eq.k10}
\ee
In this expression, $\ti g=g_1g_2/(g_1+g_2)$, and the function
$V_{Hg}(\si)$ is defined in (\ref{eq.k1}). The stationary equation
for $\ti V(\si)$ has a unique solution $\ti\si_0$, which is the
absolute minimum of this function (this is demonstrated in the same
way as for the potential (\ref{eq.k1})).
It can be shown that the quantity $\ti\si_0$
is also small in the regions where the magnetic field is weak, and
moreover
\be
\ti\si_0=\si_0(H,\ti g)+o(eH)=eH\ti g/(2\pi)+o(eH).
\label{eq.k11}
\ee
 Therefore, to estimate the magnitude of $\ti V(\ti\si_0)$, one can
 use
the relation
\be
F(x)=F(0)+F'(0_+)\cdot x+o(x), \qquad \mbox{at} \quad x\to 0_+\,,
\label{eq.k12}
\ee
where $F(x)$ is an arbitrary function differentiable at the right
from
the origin, i.e. there exists a finite limit $F'(0_+)\equiv$
$\lim_{x\to 0_+}(F(x)-F(0))/x$. Substituting $\si=\ti\si_0$
in (\ref{eq.k10}) and using equality
(\ref{eq.k12}), one can easily obtain the following value of the
function
$\ti V(\ti\si)$ at the point of the absolute minimum at $H\to 0$:
\bea
\ti V(\ti\si_0)&\approx& V_{H\ti g}(0)+\f {dV_{H\ti g}(0)}{d\si}\cdot
\ti\si_0-\f {N\ti\si^2_0}{2\ti g}\approx \nn \\
&\approx&V_{H\ti g}(0)-\f {3N(eH)^2\ti g}{8\pi^2}.
\label{eq.k13}
\eea
The second relation in this expression can be obtained with regard to
formulas (\ref{eq.k100}) and (\ref{eq.k11}). In the
same way, one can estimate the quantities in
formula (\ref{eq.k9}):
\be
V_{Hg_i}(\si_0(H,g_i))\approx V_{Hg_i}(0)-\f{N(eH)^2g_i}{4\pi^2}.
\label{eq.k14}
\ee
Comparing expressions (\ref{eq.k13}) and (\ref{eq.k14}), one can
easily
conclude that, in the set $\si_{1,2}\geq 0$, the potential
(\ref{eq.po})
can
not reach its minimum on the straight line $L$. Therefore, for
$g_2>g_1$
the
global minimum of the potential is at the point $(0,\si_{20})$, and
for
$g_1>g_2$ it is at the point $(\si_{10},0)$.

Thus, when an arbitrarily weak external magnetic field is applied at
the $P\times \Ga_5$ symmetric A-phase of the model (for which two
parameters $g_{1,2}$ are positive), ground state of this phase is
destroyed.
For $g_2>g_1$, the vacuum maintains $\Ga_5$ invariance, and phase A
transforms
to B phase of the theory. At $g_1>g_2$ the ground state is only $P$
symmetric,
and phase C spontaneously appears instead of phase A. As far as
$\si_0(H,g)\to 0$ at $H\to 0$, in both cases, magnetic field induces
the continuous second order phase transitions from phase A to
phases B or C correspondingly.

Unfortunately, for the magnetic field such that $eHg_i^2\sim 1$
or $eHg_i^2\gg~1$, we were unable to prove this fact by means of the
analytical technique. However, numerical calculations performed for a
rather wide interval of
values of $H$ confirm the above conclusion. For example, Fig.~2
depicts the
function $V_{\rm eff}(\si_1,\si_2)$ at $H=40e^3$,
$g_1=e^{-2}$, $g_2=20e^{-2}$ (here, evidently
$eHg_i^2\gg 1$). In this case, the global minimum of the potential
$\si_1=0$, $\si_2=2.77e^2$ corresponds to chiral symmetric phase of
the
model (parity is spontaneously broken).

It should also be remarked that an external magnetic field acting
upon
phases
B and C (see Fig.~1) makes their ground states more stable. The fact
is
that
there exists a rather strong fermion-antifermion attraction even at
$H=0$
(at
least one of the bare coupling constants $G_i$ is greater than
$G_c$),
leading to formation of one of the condensates $\bar\psi\psi$ or
$\bar\psi\tau\psi$, i.e. to spontaneous breaking of the initial
symmetry.
An external magnetic field acts in favor of further increasing the
condensate values, i.e. it
stabilizes these phases. Since in these cases the magnetic field is
not the
cause (catalyzer) of the spontaneous symmetry breaking (a superstrong
attraction of fermions and antifermions being the actual cause),
details of these considerations are omitted. For completeness,
however,
the phase portrait of the model at arbitrary nonzero values of the
external magnetic field and coupling constants $g_{1,2}$ is depicted
in
Fig.~3.

\section{Conclusions}

In the present paper, the influence of an external magnetic field on
the
3-dimensional
$P\ts\Ga_5$ symmetric model (\ref{eq.1}) is considered in the case,
when
bare coupling constants $G_{1,2}<G_c$ take sufficiently small values
(positive values of constants
$g_{1,2}$). The initial symmetry is
unbroken, if $H=0$. Arbitrary small external magnetic field was shown
to
induce
spontaneous breaking of $P\ts\Ga_5$ symmetry (phenomenon of magnetic
catalysis). Moreover it turned out, that $P$ symmetry of the model is
broken, while chiral symmetry remains intact, if $g_2>g_1$ (this is
phase B of the theory). When $g_1>g_2$, the external magnetic field
induces the second order phase transition from phase A to phase C,
whose vacuum is $P$ symmetric, though having no chiral invariance.

\section{Acknowledgments}

The authors are grateful to D.Ebert for useful discussions.
This work was partially supported by RFBR under grant number
98-02-16690,
and by DFG, project DFG 436 RUS 113/477.

\subsubsection*{\rightline{\ul{Appendix}}}
\subsubsection*{Algebra of $\Gamma$-matrices for the 3-dimensional
Lorentz group}

Two-component Dirac spinors realize an irreducible 2-dimensional
representation of the Lorentz group transformations of the
3-dimensional space-time. In this case, the $2\times 2$ matrices
$\gamma$ have the form
$$
 \gamma^0=\sigma_3=\left( \ba{cc} 1&0 \\ 0&-1 \ea\right),\;
 \gamma^1=i\sigma_1=\left(\ba{cc} 0&i \\ i&0 \ea\right),\;
 \gamma^2=i\sigma_2= \left(\ba{cc} 0&1 \\ -1&0 \ea\right).
\eqno \rm (A.1)
$$
These matrices have the properties
$$
 \op{Tr}(\gamma^{\mu}\gamma^{\nu})=2g^{\mu\nu};~
 ~[\gamma^{\mu},\gamma^{\nu}]=-2i\ve^{\mu\nu\al}\gamma_{\al};~~
 \gamma^{\mu}\gamma^{\nu}=-i\ve^{\mu\nu\al}\gamma_{\al}+g^{\mu\nu},
\eqno \rm(A.2)
$$
where $g^{\mu\nu}=g_{\mu\nu}=\op{diag}(1,-1,-1),$
$\gamma_{\al}=g_{\al\beta}\gamma^{\beta},$ $\ve^{012}=1$.
Moreover,
$$
 \op{Tr}(\gamma^{\mu}\gamma^{\nu}\gamma^{\al})=-2i\ve^{\mu\nu\al}.
\eqno \rm(A.3)
$$

Recently, one frequently exploits a reducible 4-dimensional
representation of the Lorentz group, and the spinors $\psi_k$ from
(1)
transform according to this representation:
$$
 \psi= \left( \ba{cc} \psi_{1}\\ \psi_{2} \ea\right).
\eqno \rm(A.4)
$$
In (A.4) $\psi_{1},\psi_{2}$ are two-component spinors. The gamma
matrices that correspond to this representation have the form
$\Gamma^{\mu}=\op{diag}(\gamma^{\mu},-\gamma^{\mu})$, where
$\gamma^{\mu}$
are given in (A.1). It is easily demonstrated that ($\mu,\nu=0,1,2$):
$$
 \op{Tr}(\Gamma^{\mu}\Gamma^{\nu})=4g^{\mu\nu};~~
 \Gamma^{\mu}\Gamma^{\nu}=\sigma^{\mu\nu}+g^{\mu\nu};~~
$$
$$
 \sigma^{\mu\nu}=\frac{1}{2}[\Gamma^{\mu},
 \Gamma^{\nu}]=\op{diag}(-i\ve^{\mu\nu\al}\gamma_{\al},
 -i\ve^{\mu\nu\al}\gamma_{\al}).
\eqno \rm(A.5)
$$
The dimensionality of the matrix algebra that acts in the
four-dimensional spinor space is equal to 16, and its generators are
$\Gamma^{\mu}$ $(\mu=0,1,2)$ and the matrix
$\Gamma^3$ (anticommuting with them):
$$
 \Gamma^3= \left( \ba{cc} 0&I \\ I&0 \ea\right),\;
\eqno \rm(A.6)
$$
where $I$ is a unit $2\times 2$ matrix. There exists one more matrix
$\Gamma^5$, which anticommutes with all matrices
$\Gamma^{\mu}$ and $\Gamma^3$:
$$
 \Gamma^5=i\left( \ba{cc} 0&-I \\ I&0 \ea \right).
\eqno \rm(A.7)
$$
We note that matrix $\tau$ entering (\ref{eq.1}) has the form
$$
 \tau= \left( \ba{cc} I&0 \\ 0&-I \ea \right).
\eqno \rm(A.7)
$$
In the space of four-dimensional spinors (A.4), one can introduce
the parity ($P$) and chiral $\Gamma_5$ transformations, those
in terms of two-component spinors $\psi_{1,2}$ look like
$$
 P:~\psi_{1k}(t,x,y)\longrightarrow\gamma^1\psi_{2k}(t,-x,y),
 ~~~\psi_{2k}(t,x,y)\longrightarrow\gamma^1\psi_{1k}(t,-x,y),
$$
$$
 \Gamma_5:~\psi_{1k}(t,x,y)\longrightarrow -i\psi_{2k}(t,x,y),
~~~\psi_{2k}(t,x,y)\longrightarrow i\psi_{1k}(t,x,y).
\eqno \rm(A.8)
$$
Discrete transformations (A.8) can be easily rewritten with the use
of
the four-component spinors (A.4)
$$
 P:~\psi(t,x,y)\ra
 i\Gamma^1\Gamma^5\psi(t,-x,y);~~~~\Gamma_5:~\psi\ra\Gamma^5\psi.
\eqno \rm(A.9)
$$

\newpage
\pagestyle{empty}

\begin{figure}
\unitlength=1mm
\special{em:linewidth 1pt}
\begin{picture}(152.67,150.33)
\linethickness{0.4pt}
\put(70.67,80.00){\vector(1,0){72.00}}
\put(80.33,69.67){\vector(0,1){55.67}}
\put(80.33,80.00){\line(-1,0){57.00}}
\put(80.33,80.00){\line(0,-1){40.67}}
\linethickness{1pt}
\put(80.33,80.00){\line(1,0){55.33}}
\put(80.33,80.00){\line(0,1){40.67}}
\put(80.33,80.00){\line(-1,0){52.00}}
\put(80.33,80.00){\line(0,-1){35.67}}
\put(80.33,80.00){\special{em:point 1}}
\put(45.00,44.66){\special{em:point 2}}
\special{em:line 1,2,1pt}
\put(75.67,121.67){\makebox(0,0)[cc]{$g_2$}}
\put(136.67,84.67){\makebox(0,0)[cc]{$g_1$}}
\put(110.33,105.33){\makebox(0,0)[cc]{Phase A}}
\put(110.33,95.33){\makebox(0,0)[cc]{$\langle\sigma_1\rangle=0,
~\langle\sigma_2\rangle=0$}}
\put(50.00,70.67){\makebox(0,0)[cc]{Phase B}}
\put(110.00,70.67){\makebox(0,0)[cc]{Phase B}}
\put(110.00,60.67){\makebox(0,0)[cc]{$\langle\sigma_1\rangle=0,
~\langle\sigma_2\rangle\not=0$}}
\put(65.67,50.33){\makebox(0,0)[cc]{Phase C}}
\put(50.67,105.33){\makebox(0,0)[cc]{Phase C}}
\put(50.67,95.33){\makebox(0,0)[cc]{$\langle\sigma_1\rangle\not=0,
~\langle\sigma_2\rangle=0$}}
\put(49.00,55.33){\makebox(0,0)[cc]{$l$}}
\end{picture}
\caption{Phase portrait of the model in terms of $g_1$ and $g_2$
at $H=0$. Here $l=\{(g_1,g_2)~:~g_1=g_2\}.$}
\end{figure}
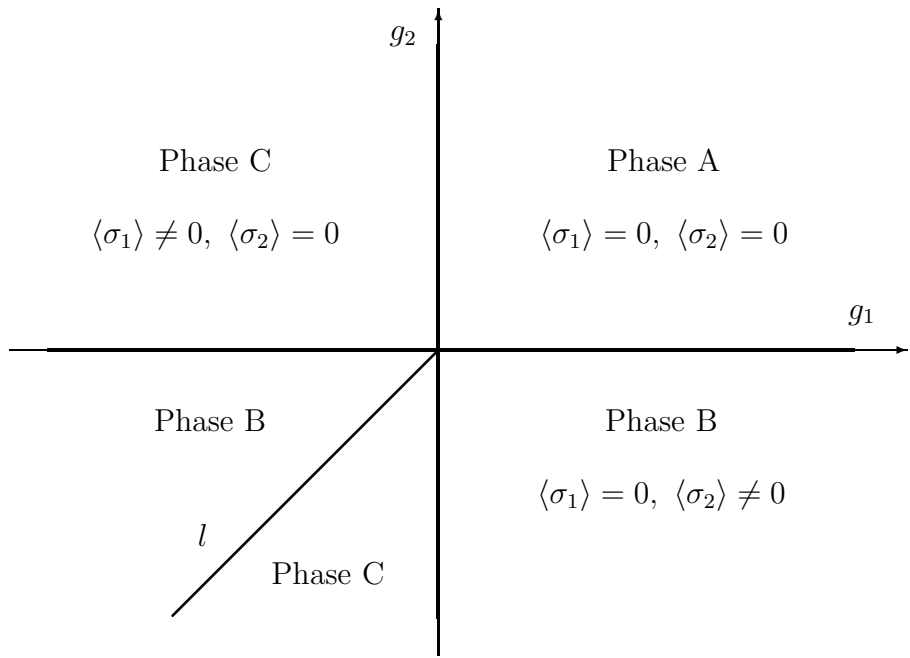

\newpage
\begin{figure}
\epsfxsize=150mm \epsfysize=116mm
\raisebox{-15mm}[0mm][0mm]{\hspace*{10mm}
 $\check V_{\rm eff}(\check\si_1,\check\si_2)$}\\
\centerline{\epsfbox{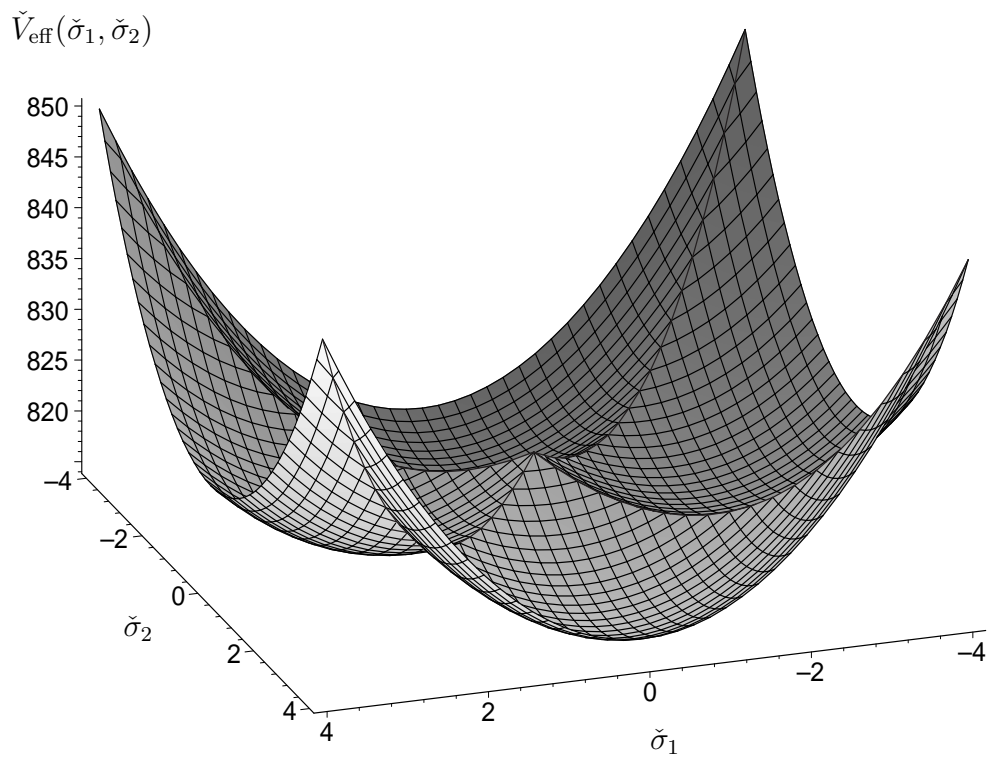}}\\
\raisebox{27mm}[0mm][0mm]{\hspace*{25mm} $\check\si_2$}\\
\raisebox{17mm}[0mm][0mm]{\hspace*{95mm} $\check\si_1$}
\caption{Diagram of the function $\check V_{\rm eff}(\check\si_1,\check\si_2)=
 V_{\rm eff}(\si_1,\si_2)/e^6$ at $H=40e^3$, $g_1=e^{-2}$, $g_2=20e^{-2}$.
Here $\check\si_1=\si_1/e^2$, $\check\si_2=\si_2/e^2$.}
\end{figure}

\newpage
\begin{figure}
\unitlength=1mm
\special{em:linewidth 1pt}
\linethickness{0.4pt}
\begin{picture}(152.67,150.33)
\put(70.67,80.00){\vector(1,0){72.00}}
\put(80.33,69.67){\vector(0,1){55.67}}
\put(80.33,80.00){\line(-1,0){57.00}}
\put(80.33,80.00){\line(0,-1){40.67}}
\linethickness{1.0pt}
\put(80.33,80.00){\line(1,0){55.33}}
\put(80.33,80.00){\line(0,1){40.67}}
\put(80.33,80.00){\line(-1,0){52.00}}
\put(80.33,80.00){\line(0,-1){35.67}}
\put(115.66,115.33){\special{em:point 1}}
\put(45.00,44.66){\special{em:point 2}}
\special{em:line 1,2,1pt}
\put(75.67,121.67){\makebox(0,0)[cc]{$g_2$}}
\put(136.67,84.67){\makebox(0,0)[cc]{$g_1$}}
\put(90.33,105.33){\makebox(0,0)[cc]{Phase B}}
\put(110.33,90.33){\makebox(0,0)[cc]{Phase C}}
\put(50.00,70.67){\makebox(0,0)[cc]{Phase B}}
\put(104.00,63.67){\makebox(0,0)[cc]{Phase B}}
\put(67.67,53.33){\makebox(0,0)[cc]{Phase C}}
\put(56.62,97.33){\makebox(0,0)[cc]{Phase C}}
\put(49.00,55.33){\makebox(0,0)[cc]{$l$}}
\end{picture}
\caption{Phase portrait of the model in terms of $g_1$ and $g_2$
at $H\not=0$. Line $l$ defined on Fig.~1.}
\end{figure}
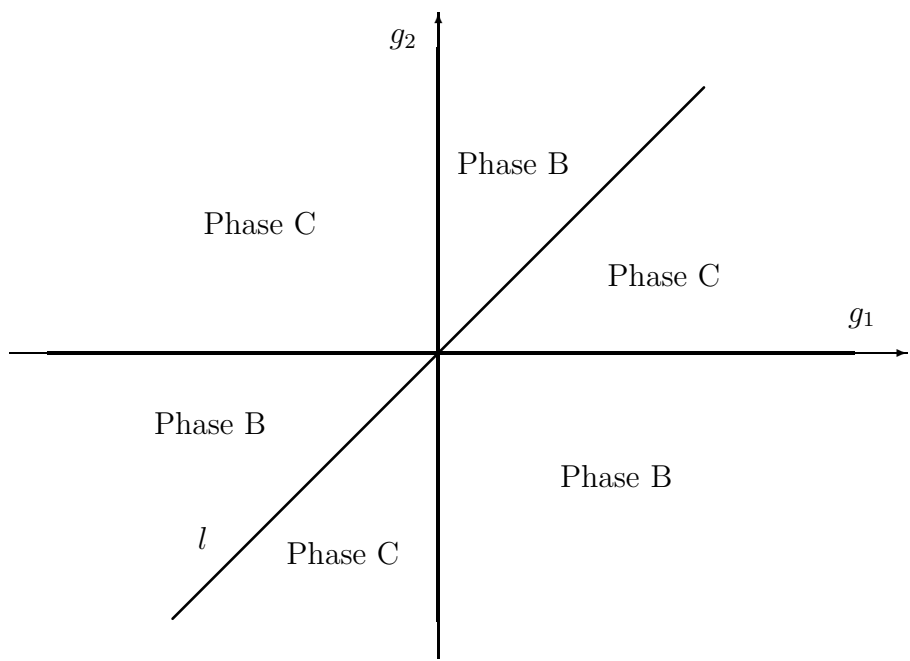

\end{document}